\documentclass[referee]{raa}           
\usepackage{graphicx,times}
\usepackage{natbib}
\usepackage{amssymb,amsmath}
\usepackage{epstopdf}
\usepackage{subcaption}
\bibpunct{(}{)}{;}{a}{}{,}

\usepackage[driverfallback=dvipdfm,pagebackref=true]{hyperref}
\hypersetup{pdftitle = The title of my PDF, pdfauthor = My name, pdfsubject= The subject, pdfkeywords = keyword1 keyword2 keyword3} 
\hypersetup{colorlinks = true, linkcolor = green, anchorcolor = red, citecolor = blue, filecolor = red, pagecolor = red, urlcolor = red}

\begin{document}

   \title{Spectroscopic study of Be-shell stars: 4 Her and 88 Her
}

 \volnopage{ {\bf 2012} Vol.\ {\bf X} No. {\bf XX}, 000--000}
   \setcounter{page}{1}

   \author{Shruthi S Bhat \inst{1}, Paul K T\inst{1}, Annapurni Subramaniam\inst{2}, Blesson Mathew\inst{3}
   }

   \institute{ Department of Physics, Christ University, Bangalore 560029, 
India; {\it shruthibhat89@gmail.com}\\
        \and
             Indian Institute of Astrophysics, II Block, Kormangala, Bangalore 560034, India\\
	\and
	   Department of Astronomy \& Astrophysics, Tata Institute of Fundamental
Research, Homi Bhabha Road, Mumbai 400005, India\\     
     %\vs \no
%   {\small Received 2012 June 12; accepted 2012 July 27}
}

\abstract{We present the optical spectroscopic study based on 41 spectra of 4 Her and 32 spectra of 88 Her, obtained over a period of 6 months. We estimate the rotational velocity of these stars from He{\sc i} lines in the blue spectral region (4000--4500 \AA). We find that these stars are likely to be rotating at a fractional critical rotation of $\sim$ 0.80. We measure the average \( I_p/I_c\) ratio to quantify the strength of the H$\alpha$ line and obtain 1.63 for 4 Her and 2.06 for 88 Her. The radius of the H$\alpha$ emission region is estimated to be  \( R_d/R_*\) $\sim$ 5.0, assuming a Keplerian disk. These stars are thus found to be fast rotators with relatively small H$\alpha$ emission region. We detect V/R variation of H$\alpha$ spectral line during the observed period. We re-estimate the periods for both the stars and obtain two periods of $\sim$ 46 days and its harmonic of 23.095 days for 4 Her and $\sim$ 86 days for 88 Her. As these two are shell stars with binaries and have low H$\alpha$ EW with the emission region closer to the central star, the V/R variation and a change in period may be an effect of the binary on the circumstellar disk.
}

   \authorrunning{Shruthi et al. }            %author_head in even pages
   \titlerunning{Spectroscopic study of Be-shell stars}  % title_head in odd pages
   \maketitle

%________________________________________________ sections below

%

\section{Introduction}
\label{Sect1}

A Classical Be star is defined as a non-supergiant B-type star whose spectrum has, or had at sometime, one or more Balmer lines in emission \citep{collins1987}. The emission lines originate from the geometrically thin, circumstellar disk rotating with near-Keplerian velocity surrounding the central star \citep{carciofi2006}. This disk is said to be formed from the material ejected from the fast-spinning central star. 

Be-shell stars are ordinary Be stars seen edge-on, so that the line of sight towards the star probes the circumstellar, equatorial disk \citep{porter2003}. They have sharp and deep absorption components in the centres of double-peaked emission lines. If the absorption in between the two peaks reaches below the continuum level, it is called a shell profile. Shell stars also have a strongly rotationally broadened photospheric lines and additional narrow absorption lines \citep{rivinius2006}. A significant fraction of Be stars, and thus also shell stars, undergoes long-term variability of V/R which is an indication of the Keplerian motion in the circumstellar disk \citep{hanuschik1995}. One-third of all double-peaked profiles exhibit what is called violet-to-red emission peak height ratio (V/R) variations \citep{catanzaro2013}. It is one of the main characteristics describing the double-peaked emission lines of Be stars.

In this paper we mainly present the V/R variability of two such shell stars i.e., 4 Her and 88 Her (see Table~\ref{Tab1}) based on spectroscopic data. We have obtained 41 spectra of 4 Her and 32 spectra of 88 Her over a period of about six months. We discuss the observed features and changes in the spectra of these stars. We have estimated the radius of the circumstellar disk using the H$\alpha$ line. We also determined the rotational velocity of the central star from the He{\sc i} lines.

The paper is arranged as follows. The following section gives a brief overview of these two stars mainly on there spectroscopic variability from previous studies. Section 3 addresses the details of spectral
observations and data reduction techniques. In section 4, we present the spectra and discuss the major results from the spectral line analysis of both the stars. The conclusions drawn from this study are listed in section 5.

\begin{table}
\caption{Program stars.}
\centering
\begin{tabular}{cccccccc}
\hline
\hline
$\bf HD$ & $\bf HR$ & $\bf Name$ & $\bf Other$ & $\bf Spectral$ & $\bf RA$ & $\bf \delta$ & $\bf V$\\
 &  &  & $\bf names$ & $\bf Type$ &  \\ 
\hline
142926 & 5938 & 4 Her &	V839 Her & B7 IVe shell	& 15 05 30.59216 & +42 33 58.2934 & 5.75\\
162732 & 6664 &	88 Her & V744 Her, $\zeta$ Her& B6 IVe & 17 50 03.33579 & +48 23 38.9598 & 6.89\\
\hline
\end{tabular}
\label{Tab1}
\end{table}

\section{Previous studies}

\subsection{4 Her}
\label{Sect2.1}
4 Her is a well known star and rather frequently studied Be-shell star. It was first recognized as a Be star by \cite{heard1939} and \cite{mohler1940}. \cite{hubert1971} reported remarkable spectral changes of the star which occurred between 1953 and 1970. \cite{koubsky1994} identified two different periods of 28 and 43 years when the H$\alpha$ emission was absent. \cite{harmanec1973} suggested the object is a single-line spectroscopic binary.

	\cite{koubsky1997} showed that H$\alpha$ V/R and radial velocities of the shell lines followed a 46.18 day period. This was also confirmed by \cite{rivinius2006}, who also reported changes in the equivalent width. \cite{stefl2007} reported the star to show orbital phase-locked variations and the variations to be coherent over more than 80 cycles. It also exhibits central quasi emission bumps (CQEB) which is the apparent doubling of some shell lines and was first reported by \cite{koubsky1997}. The shell lines were reported to be typically broad for all CQE stars by \cite{rivinius1999}. They report that further investigation of the period as well as the nature of its secondary component has to be carried out. \cite{catanzaro2013} estimated the radius of the circumstellar disk to be 4.3 \( R_*\). 
	
\subsection{88 Her}
\label{Sect2.2}
88 Her was first discovered in 1959 to be a Be-shell star by \cite{bidelman1960}. They described the H$\alpha$ to show a very strong double emission line,  with an intense absorption core almost centrally dividing the broad emission into two very nearly equal components during 1959. \cite{harmanec1974} improved their previous estimation of period and gave a much more precise value of the period as 86.59d and also concluded it to be a spectroscopic binary.
	
	\cite{hirata1978} compared Pleione to 88 Her and saw a shell phase of 88 Her again in 1978 after a decline of variation in 1977. \cite{doazan1982} refined the period of 88 Her to be 86.7221 days which is followed by the periodic variations of the shell Balmer lines as well as the V/R ratio of the double-peaked H$\alpha$ emission line. \cite{barylak1986} discussed the variations from 1972 -- 1983,  during which 88 Her underwent four epochs of variation: a) 1967 -- 1972 : Decreasing Be-shell phase where shell lines disappeared and emission weakened; b) 1972 -- 1977 : Quasi-normal B phase where the star showed almost constant behaviour with very mild hydrogen shell phase; c) 1977 -- 1978 : Be-shell phase where the reappearance of the shell metallic lines were seen; d) 1978 -- 1983 : Strong Be-shell phase where shell spectrum developed strongly both in the H$\alpha$ emission and in the shell absorption line strength. \cite{hirata1995} observed decreasing metallic shell and mild Be phase between 1984 -- 1986, a recovery phase from 1987 -- 1991 and later reported the star to be in mild Be phase between 1992 -- 1993. \cite{mennickent1991} gave the circumstellar disk radius measured from H$\alpha$ as 1.14 \( R_*\). They also confirmed the period of 86.7221 days using observations before 1989. 
	
\section{Observations and data reduction}
\label{Sect3}

\begin{table}[!htb]
\centering
\caption{Journal of observations for 4 Her and 88 Her.}
\begin{tabular}{ccc|c}
\hline
\hline
$\bf Date\ of$ & $\bf Spectral$ & \multicolumn{2}{c}{$\bf No.\ of\ spectra$ }\\
$\bf Observation$ & $\bf Range$ & \multicolumn{2}{c}{$\bf (Integration\ time$ }\\
$\bf in\ 2009$ & $\bf $\AA$ $ & \multicolumn{2}{c}{$\bf in\ seconds)$ }\\
\hline
 &  & $\bf 4\ Her$ & $\bf 88\ Her$\\
\hline
3 February & 6200 -- 6800 & 1 (1800) & ...\\
4 February & 6200 -- 6800 & 1 (2400) & ...\\
6 April & 4000 -- 4600 & 1 (2400) & ...\\ 
14 April & 4000 -- 4600 & 1 (2700) & 1 (2700)\\
26 April & 6200 -- 6800 & 2 (2700) & 2 (2700)\\
27 April & 6200 -- 6800 & 2 (2700) & 1 (2700)\\
28 April & 6200 -- 6800 & 2 (1800, 2700) & 2 (2700)\\
29 April & 6200 -- 6800 & 2 (1800) & 2 (2700)\\
30 April & 6200 -- 6800 & 2 (1800) & 2 (2400)\\
1 May & 6200 -- 6800 & 2 (1800) & 3 (1800, 2400, 2400)\\ 
2 May & 4000 -- 4600 & 2 (1800) & 1 (1800)\\ 
3 May & 4000 -- 4600 & 2 (1800) & 2 (2400)\\
5 May & 4000 -- 4600 & 2 (1800) & 2 (2400)\\
6 May & 4000 -- 4600 & 2 (1800) & 2 (2400)\\
10 May & 4000 -- 4600 & 2 (2700) & ...\\
11 May & 4000 -- 4600 & 2 (2700) & 2 (2700)\\
13 May & 4000 -- 4600 & 3 (2700) & 1 (2700)\\
2 June & 3800 -- 4300 & 1 (2400) & 1 (2700)\\
16 June & 3800 -- 4300 & 1 (1800) & 1 (2400)\\
23 June & 3800 -- 4300 & 2 (2400) & 2 (2400)\\
25 June & 3800 -- 4300 & ... & 2 (2400)\\
11 July & 6200 -- 6800 & 2 (2400) & 2 (2700)\\
22 July & 6200 -- 6800 & 2 (2700) & ... \\
23 July & 6200 -- 6800 & 2 (2700) & ... \\
27 July & 6200 -- 6800 & ... & 1 (2700)\\
\hline
\label{Tab2}
\end{tabular}
\end{table}

%\begin{table}[!htb]
%\centering
%\caption{Journal of observations from BeSS}
%\begin{tabular}{ccccccc}
%\hline
%\hline
%$\bf Star$ & $\bf Date\ of$ & $\bf Telescope$ & $\bf Instrument$ & $\bf Resolving$ & $\bf Spectral$ & $\bf Ref.$\\
%$\bf Name$ & $\bf Observation\ (2009)$ & & & $\bf power$ & $\bf range\ ($\AA$)$ & \\
%\hline
%4 Her & 6 May & Meade LX200 & LHIRES & 6000 & 6450-6850 & 1\\
% & 15 May & Schmidt-Cassegrain & LHIRES & 6000 & 6500-6700 & 2\\
%88 Her & 22 March & C11 & echelle & 10000 & 4290-7100 & 3\\
% & 12 July & Schmidt-Cassegrain & LHIRES & 6000 & 6500-6700 & 2\\
%\hline 
%\\
%\multicolumn{7}{l}{%
%  \begin{minipage}{10cm}%
%   \small References: 1 -- J.-N. Terry; 2 -- J. Guarro Fl\'{o}; 3 -- C.Buil.
%  \end{minipage}%
%}
%\end{tabular}
%\label{Tab10}
%\end{table}

%mathematics mode
The spectra of the two stars were obtained using the Universal Astronomical Grating Spectrograph (UAGS) at the Cassegrain focus of the 1.0m Carl Zeiss reflector located at Vainu Bappu Observatory, Kavalur, India \((12^\circ 34^\prime) \) and operated by the Indian Institute of Astrophysics (IIA). The CCD used for imaging consists of \(1024 \times 1024\) pixels of 24 $\mu$m size, where the central \(1024 \times 300\) pixels were used for spectroscopy and the typical readout noise is of about \(4.8e^-\) and the gain is \(1.22e^-/\)ADU. All the spectra have been acquired during several observation runs from February 2009 to July 2009 and the journal of observations is given in Table~\ref{Tab2}. The grating used for this particular observation is the Bausch and Lomb 1800 lines per millimetre grating, which in combination with the slit provided a resolution of 1~\AA ~at H$\alpha$. The medium resolution data for the blue spectral region which included absorption lines like H$\gamma$ to H$\theta$ and also He{\sc i} lines were taken in the wavelength region 3800 -- 4600~\AA ~and for the red region having H$\alpha$ in emission were taken in the range 6200 -- 6800~\AA . 

The reduction of spectra,  which included the subtraction of the bias frame,  correcting for the flat-field,  the extraction of the aperture and wavelength calibration were performed using several routines in the NOAO/IRAF~\footnote{IRAF is distributed by the National Optical Astronomy Observatory, which is operated by the Association of Universities for Research in Astronomy, Inc.}(Image Reduction and Analysis Facility) package. Dome flats were used to correct the pixel-to-pixel quantum efficiency variations. The wavelength calibration was performed using Fe-Ar arc lamp spectra. Typical S/N near H$\alpha$ for 4 Her is $\sim$160 from 20 spectra and for 88 Her is $\sim$100 from 15 spectra.

All the spectra were initially normalized to the continuum. IRAF tasks were later used to measure parameters of the emission line profiles,  such as Equivalent Widths (EW), Full Width at Half Maximum (FWHM), V/R ratios, peak separations ($\Delta$V) and \( I_p/I_c\). All the measurements are reported in the next section.

\section{Analysis and discussion}
\begin{figure}[''h'']
    \begin{subfigure}{0.5\textwidth}
    \centering
    \includegraphics[width=6.5cm,  height=6.5cm, angle=270]{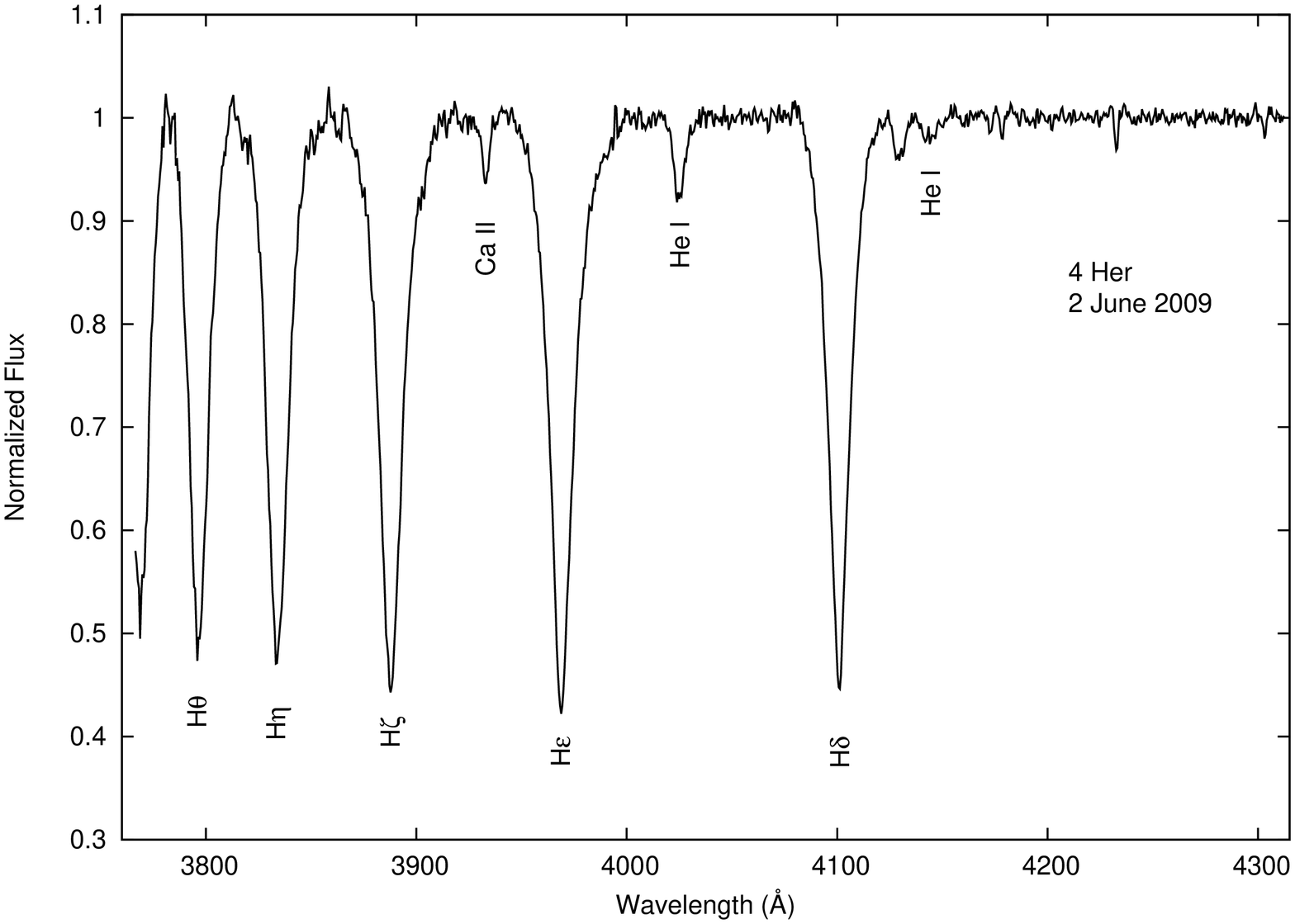}
    \end{subfigure}%
    %\hspace{0.3cm}
    \begin{subfigure}{0.5\textwidth}
    \centering
    \includegraphics[width=6.5cm,  height=6.5cm, angle=270]{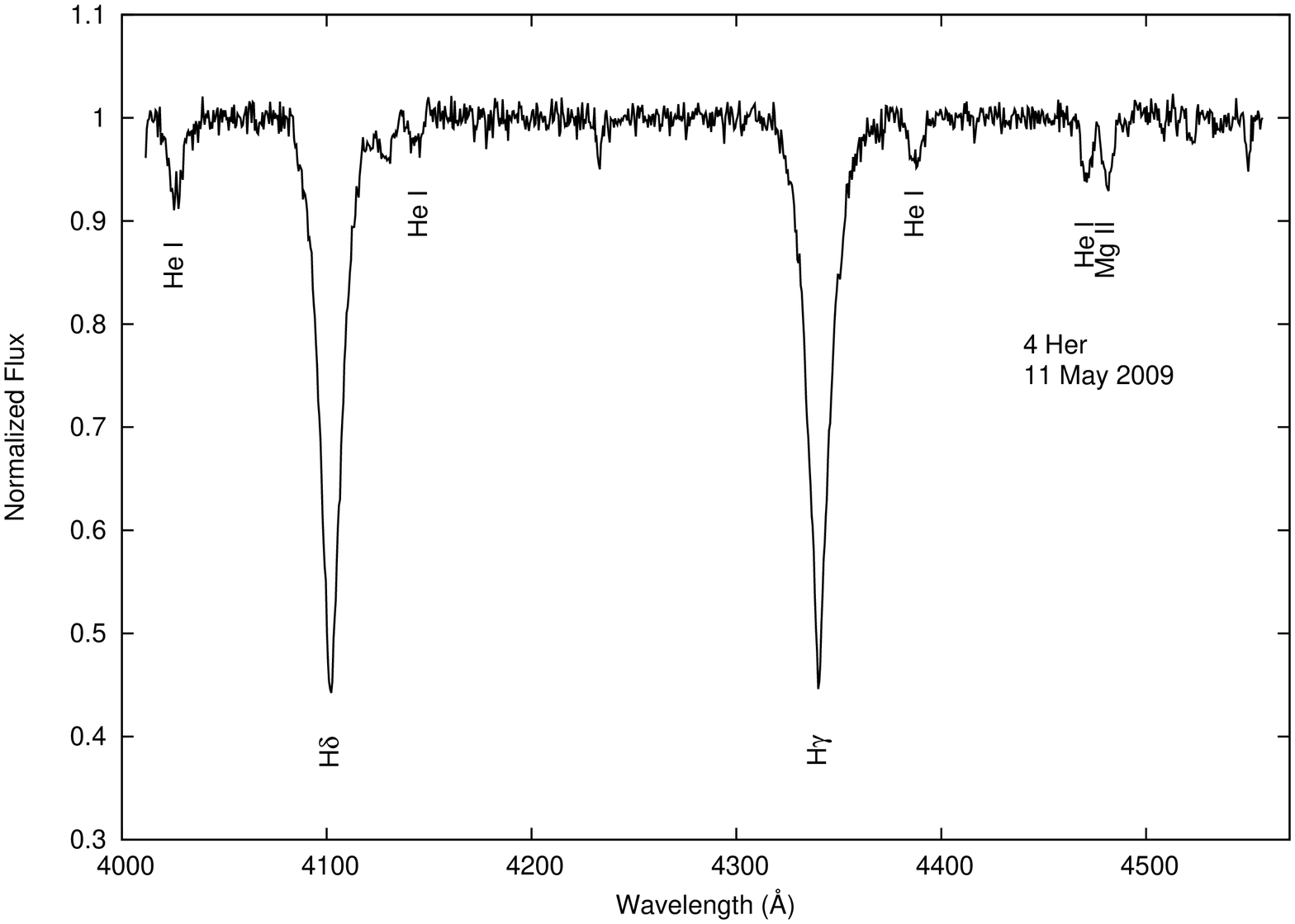}
    \end{subfigure}%
    
    \begin{subfigure}{0.5\textwidth}
    \centering
    \includegraphics[width=6.5cm,  height=6.5cm, angle=270]{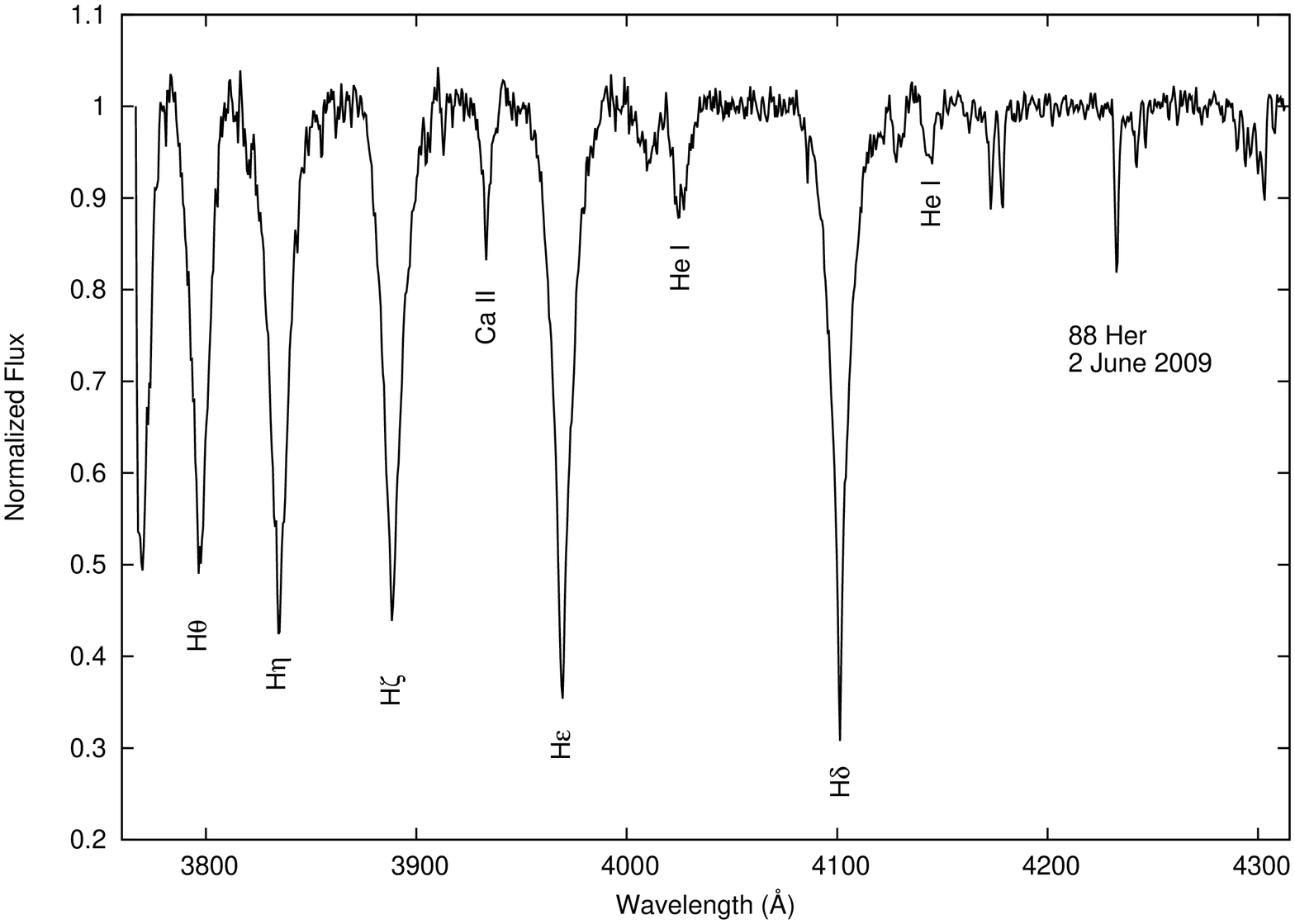}
    \end{subfigure}%
    %\hspace{0.3cm}
    \begin{subfigure}{0.5\textwidth}
    \centering
    \includegraphics[width=6.5cm,  height=6.5cm, angle=270]{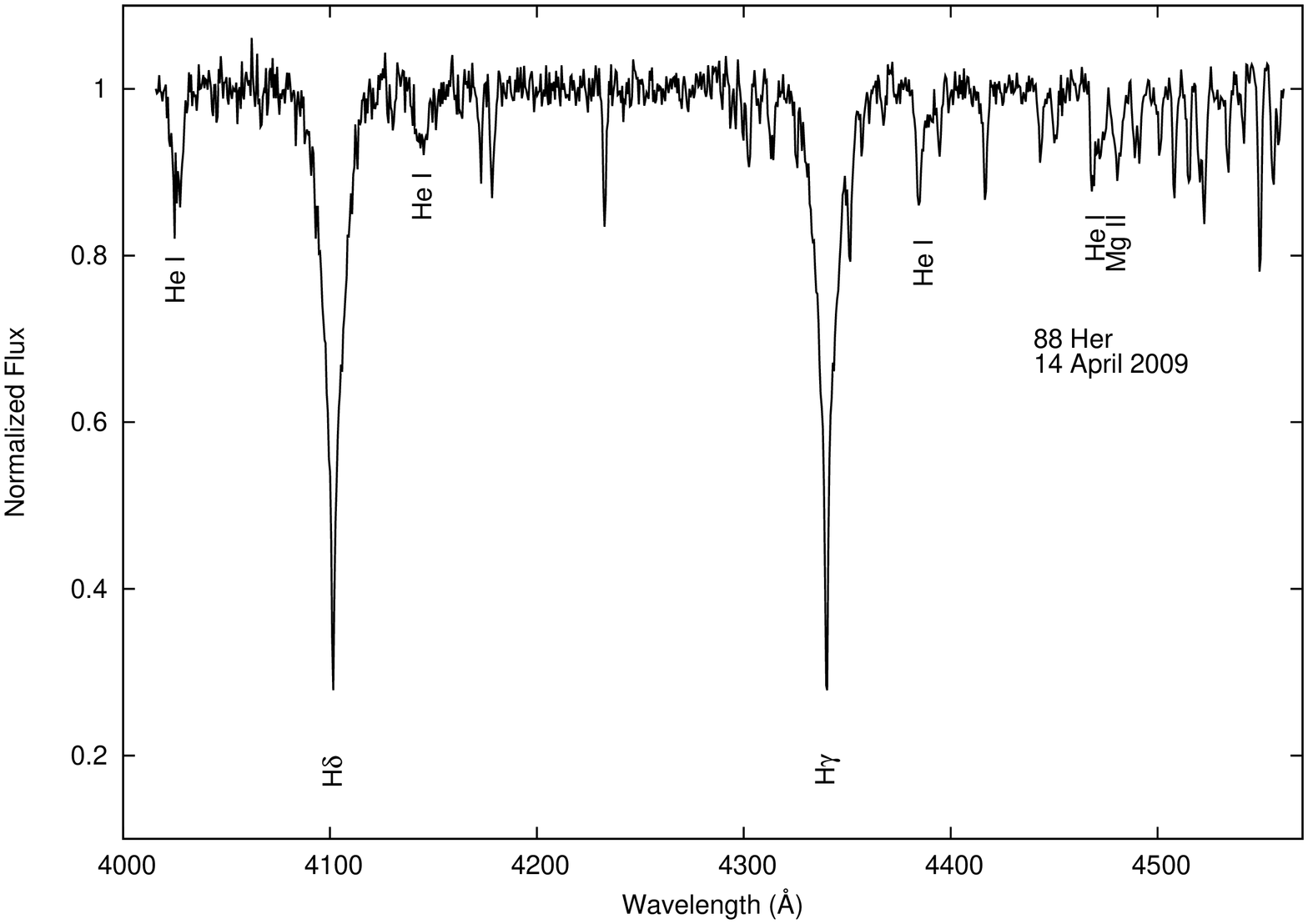}
    \end{subfigure}%
	\caption{Spectra of 4 Her (top) and 88 Her (bottom) in the wavelength region 3700 -- 4600~\AA ~along with the prominent absorption lines marked on the spectra.}
	\label{Fig1}
\end{figure}

4 Her was observed from February to July, 2009 and 88 Her was observed from April to July, 2009. Representative sample spectra for the two stars are shown in Figure \ref{Fig1} and \ref{Fig2} respectively. In the following section, we discuss the rotational velocity study of both the stars. Sect. \ref{Sect4.2} deals with the radius of the circumstellar disk and the method of its estimation. V/R variation for the stars during the observation period is separately studied in Sect. \ref{Sect4.3}.

\subsection{Rotational velocity estimation}
\label{Sect4.1}

Rotational velocity ($\textit{v}$ sin $\textit{i}$) of Be stars are in general very high compared to B-type stars. Be stars are said to be rapid rotators and current statistics indicate that for a few Be stars, rotational velocity reaches the critical velocity \citep{rivinius2013}.

%mathematics mode
%\begin{equation}
% v\ sin\ \emph{i} = \frac{c(FWHM)}{2\lambda_\circ{ln2}}
% \label{eq1} 
%\end{equation} 
For rotational velocity estimation, we have used the He{\sc i} absorption lines in the blue spectral region, He{\sc i} $\lambda$4026, 4143, 4387 and 4471 (refer Figure \ref{Fig1}). These lines are assumed not to be affected by the emission from the disk. Whereas, in the red spectral region (Figure \ref{Fig2}), He{\sc i} $\lambda$6678 absorption line is not used for the rotational velocity estimation, since there could be emission component present in this line. For 88 Her, He{\sc i} $\lambda$4387 was very weak and thus was not considered for $\textit{v}$ sin $\textit{i}$ estimation.

\cite{steele1999} derived the spectral class and rotational velocities for a sample of 58 Be stars. They made a fit to the FWHM - $\textit{v}$ sin $\textit{i}$ correlation of \cite{slettebak1975}and obtained relations Eq. 1 - 4  in their paper for the four He{\sc i} lines mentioned above. We estimated the $\textit{v}$ sin $\textit{i}$ for both 4 Her and 88 Her, using those relations between FWHM and $\textit{v}$ sin $\textit{i}$ for He{\sc i} $\lambda$4026, 4143, 4387 and 4471. The FWHM of the He{\sc i} lines and the estimated $\textit{v}$ sin $\textit{i}$ values are shown in Table~\ref{Tab3} for 4 Her and in Table~\ref{Tab4} for 88 Her. The average $\textit{v}$ sin $\textit{i}$ estimated for 4 Her and 88 Her are shown in Table~\ref{Tab5}. Out of the four He{\sc i} lines, He{\sc i} $\lambda$4026 was well resolved in most of the spectra and was consistent throughout the sample.  The error tabulated for all the values corresponds to the standard deviation.

 %\cite{slettebak1982} 
 
%\cite{catanzaro2013} 
 
%\cite{slettebak1982} 

%\cite{slettebak1976} 
 
%\cite{catanzaro2013} 
 %\cite{slettebak1976} 

\begin{table}
\centering
\caption{FWHM and \emph{v}\ sin\ \emph{i} measurements for 4 Her from the spectra using He{\sc i} $\lambda$4026, 4143, 4387 and 4471.}
\begin{tabular}{ccccccccc}
\hline
\hline
$\bf Date\ of$ & \multicolumn{2}{c}{$\bf He${\sc i}$\ \lambda4026 $} & \multicolumn{2}{c}{$\bf He${\sc i}$\ \lambda4143 $} & \multicolumn{2}{c}{$\bf He${\sc i}$\ \lambda4387 $} & \multicolumn{2}{c}{$\bf He${\sc i}$\ \lambda4471 $}\\
$\bf Observation$ & $\bf FWHM $ & $\textit{v}$ sin $\textit{i}$ & $\bf FWHM $ & $\textit{v}$ sin $\textit{i}$ & $\bf FWHM $ & $\textit{v}$ sin $\textit{i}$ & $\bf FWHM $ & $\textit{v}$ sin $\textit{i}$ \\
 & $\bf ($\AA$)$ & $(\rm km $ $ s^{-1})$ & $\bf ($\AA$)$ & $(\rm km $ $ s^{-1})$ & $\bf ($\AA$)$ & $(\rm km $ $ s^{-1})$ & $\bf ($\AA$)$ & $(\rm km $ $ s^{-1})$\\
\hline
06/04/09 & 7.4 & 338.2 &  --   &  --     & 7.8  & 326.2 & 6.2 & 254.2 \\
14/04/09 & 5.6 & 257.5 & 7.7 & 342.8 & 11.7 & 493.4 & 5.8 & 238.8 \\
02/05/09 & 6.4 & 291.3 &  --   &  --     & 7.2  & 302.1 & 6.9 & 285.4 \\
02/05/09 & 6.2 & 285.5 & --    &  --     & 8.4  & 354.1 & 5.5 & 227.1 \\
03/05/09 & 5.2 & 238.4 & 4.6 & 204.8 & 6.3  & 265.5 & 6.0 & 245.8 \\
03/05/09 & 6.6 & 300.4 &  --   &   --    & 7.8  & 329.0 & 5.9 & 245.1 \\
05/05/09 & 6.1 & 279.4 & 5.8 & 257.5 & 7.1  & 298.7 & 7.0 & 290.2 \\
05/05/09 & 6.5 & 296.9 & 8.4 & 374.2 & 7.0  & 292.5 & 7.8 & 320.1 \\
06/05/09 & 6.4 & 293.6 &  --   &  --     & 9.7  & 408.5 & 6.3 & 259.8 \\
06/05/09 & 8.3 & 379.1 &  --   &  --     & 7.8  & 327.8 & 6.6 & 272.1 \\
10/05/09 & 8.7 & 397.0 &  --   &  --     & 8.3  & 348.4 & 7.1 & 291.2 \\
10/05/09 & 8.5 & 390.7 &  --   &  --     & 7.7  & 325.6 & 7.2 & 296.5 \\
11/05/09 & 6.3 & 286.8 & 9.3 & 413.0 & 7.7  & 322.4 & 5.4 & 222.0 \\
11/05/09 & 6.1 & 277.9 & 6.9 & 305.0 & 7.3  & 307.9 & 7.2 & 299.0 \\
13/05/09 & 6.2 & 285.3 & 7.1 & 316.6 & 6.9  & 291.4 & 7.0 & 287.7 \\
13/05/09 & 7.8 & 356.6 & 7.8 & 345.6 & 7.7  & 323.8 & 6.4 & 264.9 \\
13/05/09 & 7.8 & 357.7 & 7.9 & 349.8 & 3.4  & 141.7 & 6.7 & 274.6 \\
02/06/09 & 6.2 & 285.0 & 8.8 & 392.2 &  --    &   --    &  --   &  --     \\
16/06/09 & 6.7 & 307.8 & 5.4 & 241.8 &  --    &   --    &  --   &  --     \\
23/06/09 & 6.9 & 317.2 & 7.5 & 333.2 &  --    &   --    &  --   &  --     \\
23/06/09 & 6.3 & 287.9 & 6.5 & 288.7 &  --    &   --    &  --   &  --     \\
\hline
\end{tabular}
\label{Tab3}
\end{table}

\begin{table}
\centering
\caption{FWHM and \emph{v}\ sin\ \emph{i} measurements for 88 Her from the spectra using He{\sc i} $\lambda$4026, 4143 and 4471.}
\begin{tabular}{ccccccc}
\hline
\hline
$\bf Date\ of$ & \multicolumn{2}{c}{$\bf He${\sc i}$\ \lambda4026 $} & \multicolumn{2}{c}{$\bf He${\sc i}$\ \lambda4143 $} & \multicolumn{2}{c}{$\bf He${\sc i}$\ \lambda4471 $}\\
$\bf Observation$ & $\bf FWHM $ & $\textit{v}$ sin $\textit{i}$ & $\bf FWHM $ & $\textit{v}$ sin $\textit{i}$ & $\bf FWHM $ & $\textit{v}$ sin $\textit{i}$ \\
 & $\bf ($\AA$)$ & $(\rm km $ $ s^{-1})$ & $\bf ($\AA$)$ & $(\rm km $ $ s^{-1})$ & $\bf ($\AA$)$ & $(\rm km $ $ s^{-1})$ \\
\hline
14/04/09 & 6.6 & 301.3 & 7.5  & 333.8 & 8.2 & 337.4 \\
02/05/09 & 7.2 & 328.3 & 4.2  & 184.8 & 5.9 & 244.6 \\
03/05/09 & 5.4 & 245.1 & 6.3  & 282.2 & 7.5 & 308.9 \\
03/05/09 & 7.6 & 348.7 &  --    &  --    & 7.4 & 305.3 \\
05/05/09 & 7.0 & 321.8 & 4.9  & 219.7 & 7.8 & 323.2 \\
05/05/09 & 5.1 & 232.8 &  --    &  --     & 6.6 & 270.2 \\
06/05/09 & 7.5 & 342.8 & 7.2  & 322.6 & 7.0 & 289.8 \\
06/05/09 & 7.5 & 342.0 &  --    &  --     & 6.7 & 276.8 \\
11/05/09 & 6.5 & 295.5 & 6.2  & 277.3 & 6.8 & 279.9 \\
11/05/09 & 6.6 & 300.7 & 6.3  & 280.2 & 6.2 & 254.8 \\
13/05/09 & 6.9 & 315.2 &  --    &  --     & 6.7 & 278.0 \\
02/06/09 & 9.2 & 422.1 & 5.5  & 246.1 & --   &  --     \\
16/06/09 & 7.1 & 323.8 & 6.2  & 274.7 & --   &  --     \\
23/06/09 & 5.7 & 263.2 & 11.7 & 519.4 & --   &  --     \\
23/06/09 & 9.2 & 422.6 & 7.3  & 323.1 & --   &  --     \\
25/06/09 & 6.6 & 302.2 & 8.2  & 365.6 & --   &  --     \\
25/06/09 & 5.7 & 260.7 &  --    &  -- & --   &  --     \\
\hline
\end{tabular}
\label{Tab4}
\end{table}

\begin{table}
\centering
\caption{Rotational velocity parameters. \emph{v}\ sin\ \emph{i} measured from the spectra for both the stars is compared with two other estimations; $\omega$ was calculated using \(v_c\) given in Table 2 of \cite{yudin2001}, who interpolated values by \cite{moujtahid1999}.}
\begin{tabular}{cccc}
\hline
\hline
$\bf Parameters$ &  & $\bf 4\ Her$ & $\bf 88\ Her$\\
\hline
$\textit{v}$ sin $\textit{i}$ & 4026~\AA & 310 $\pm$ 10 (21\textsuperscript{\dag}) & 316	 $\pm$ 13 (17\textsuperscript{\dag})\\
$(\rm km $ $ s^{-1})$ & 4143~\AA & 320 $\pm$ 17 (13\textsuperscript{\dag}) & 303 $\pm$ 25 (12\textsuperscript{\dag})\\
 & 4387~\AA & 321 $\pm$ 17 (17\textsuperscript{\dag}) & -- \\
 & 4471~\AA & 269 $\pm$ 7 (17\textsuperscript{\dag}) & 288 $\pm$ 9 (11\textsuperscript{\dag})\\
 & Average & 305 $\pm$ 13 & 302 $\pm$ 16\\
                              &  Slettebak & 300\textsuperscript{*} & 300\textsuperscript{**}\\
                              &  Catanzaro & 275 & 40\\
\hline                              
$v_c$ $(\rm km $ $ s^{-1})$  & Yudin & 362 & 374\\       
\hline
$\omega$ &  4026~\AA & 0.86 & 0.85\\
		 &  4143~\AA & 0.88 & 0.81\\
		 &  4387~\AA & 0.89 & --\\
		 &	4471~\AA & 0.74 & 0.77\\
		 & Average & 0.84 & 0.81\\
\hline 
\multicolumn{4}{l}{\textsuperscript{\dag} \footnotesize{Total number of spectra from which the value was averaged}}\\
\multicolumn{4}{l}{\textsuperscript{*} \footnotesize{\cite{slettebak1982},} \textsuperscript{**} \footnotesize{\cite{slettebak1976}}}\\
\end{tabular}
\label{Tab5}
\end{table}

For 4 Her, the average values obtained were compared with the values from \cite{slettebak1982} and \cite{catanzaro2013}. Similarly for 88 Her, the average $\textit{v}$ sin $\textit{i}$ obtained were compared with that of \cite{slettebak1976}. All the values and the average for both the stars seem to be consistent with that of the literature except with that in \cite{catanzaro2013} and is also found to be within the error.

Shell stars, as defined by \cite{rivinius2006} are stars with strongly rotationally broadened photospheric lines. Hence the $\textit{v}$ sin $\textit{i}$ values estimated in this study from the photospheric lines show a very high value as expected and also match well with the literature values. \cite{catanzaro2013} quotes 40$\rm km $ $ s^{-1}$ for 88 Her using Mg{\sc ii} 4481, which is underestimated compared to the values we have obtained. 

We have estimated the critical fractional rotation velocity of these stars, as these are expected to rotate close to the critical or break-up velocity. The critical velocity, \( v_c \) value was taken from \cite{yudin2001} which was estimated for a particular spectral class and luminosity class. Thus, for the spectral type indicated in Table~\ref{Tab1}, \( v_c \) was obtained and critical fractional rotation, $\omega$ given by \( v\ sin\ \emph{i}/{v_c} \) was estimated. The uncertainty in $\omega$ is not only from $\textit{v}$ sin $\textit{i}$ but also from the spectral and the luminosity classes \citep{rivinius2006}. The critical fractional rotation obtained in our study using the He{\sc i} lines were compared with the values obtained by \cite{rivinius2006}.   %\cite{rivinius2006} 
They estimated an average critical fractional rotation, $\bar{\omega}$ to be 81 $\pm$ 12\%, for shell stars, by considering 27 shell stars. We find that our estimation is similar to the literature value.

\subsection{Disk radius estimation}
\label{Sect4.2}

\begin{figure}
	\begin{subfigure}{0.5\textwidth}
    \centering
    \includegraphics[width=6.5cm,  height=7cm, angle=270]{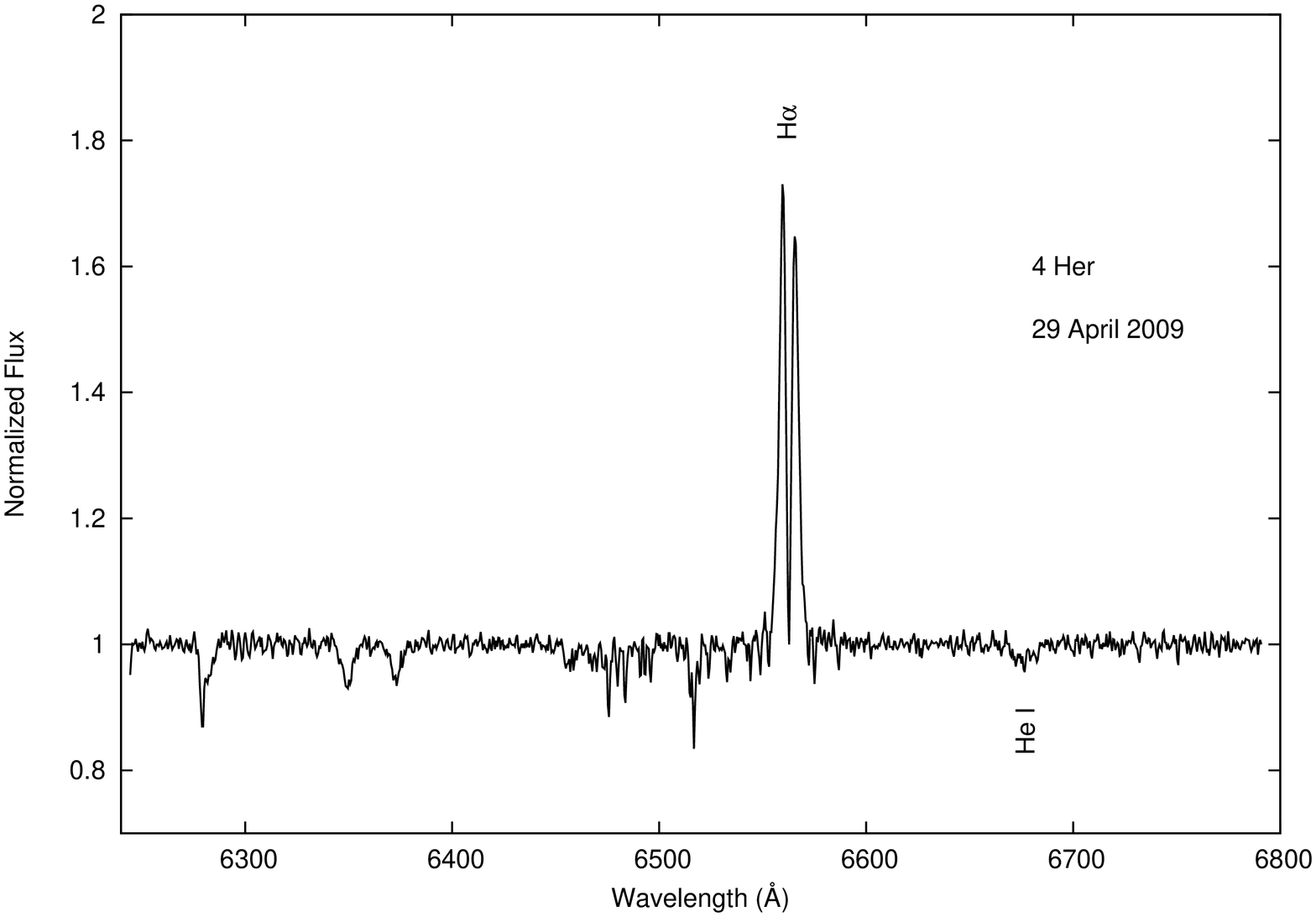}
    \end{subfigure}%
    %\hspace{0.3cm}
    \begin{subfigure}{0.5\textwidth}
    \centering
    \includegraphics[width=6.5cm,  height=7cm, angle=270]{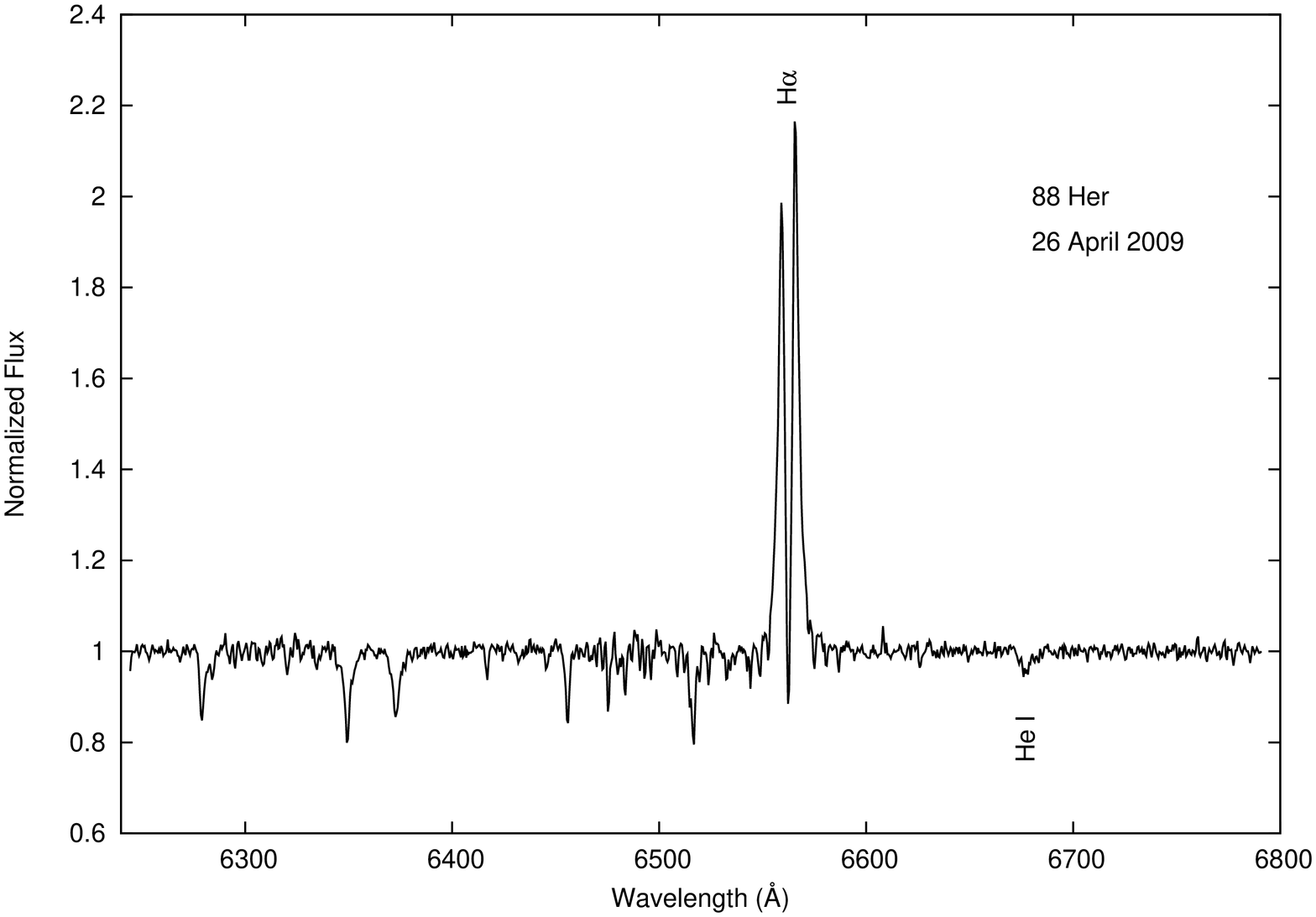}
    \end{subfigure}%
\caption{Spectra of 4 Her (left) and 88 Her (right) in the wavelength region 6200 -- 6800~\AA ~showing H$\alpha$ double-peaked emission along with He{\sc i} absorption line.}
\label{Fig2}
\end{figure}

\begin{figure}
    \centering
    \includegraphics[width=13cm,  height=9cm]{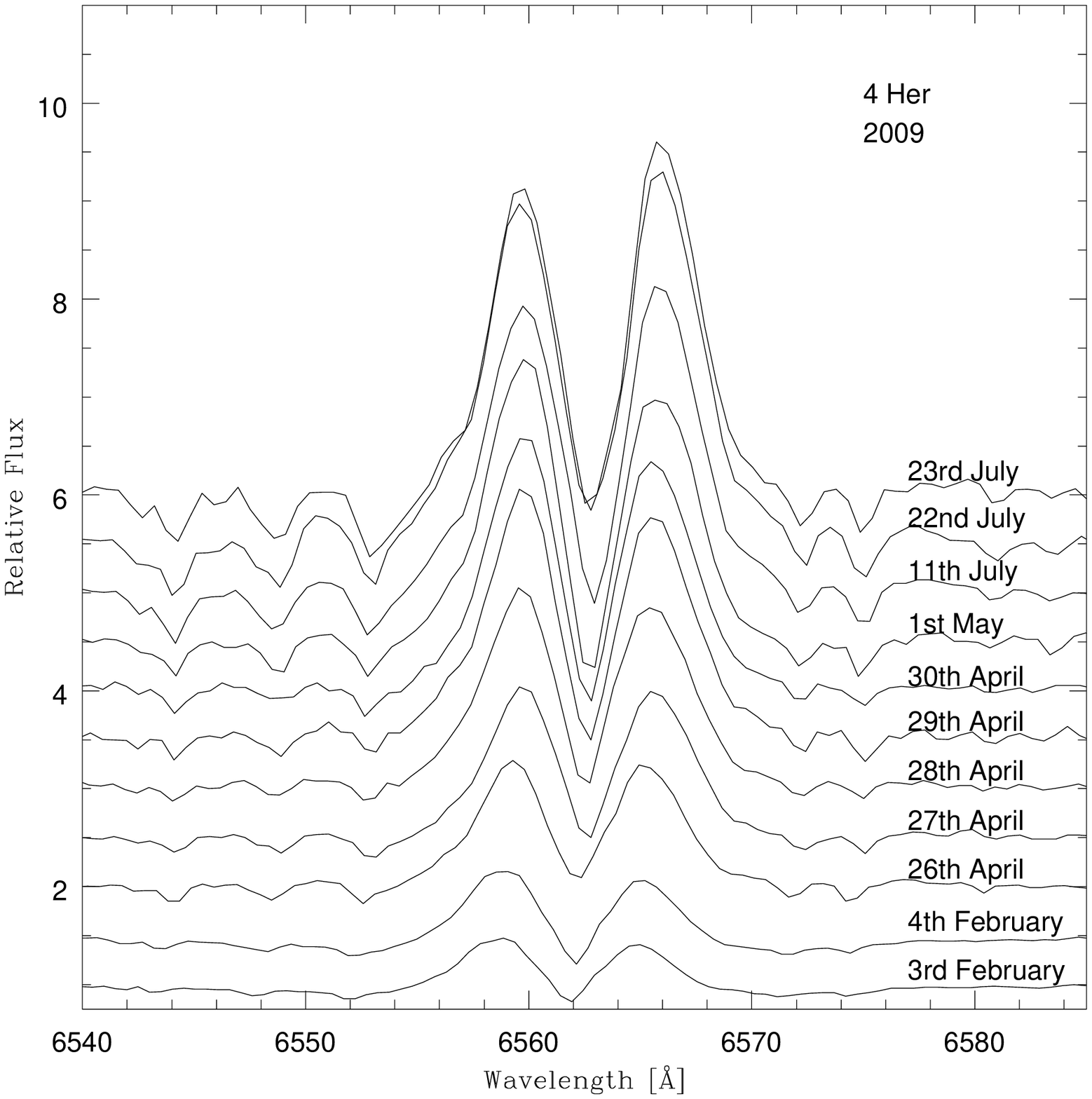}
    \includegraphics[width=13cm,  height=9cm]{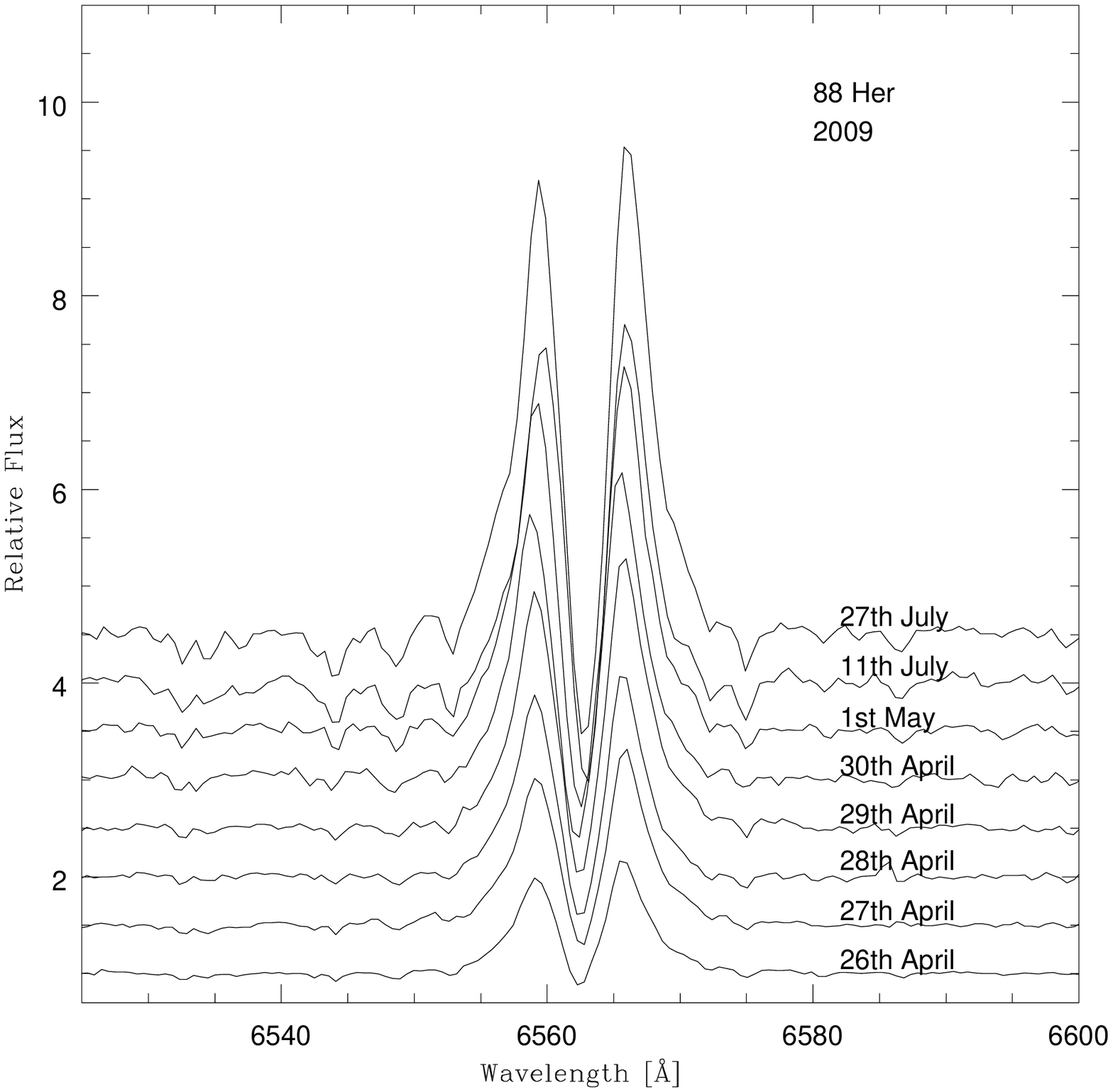}
    \caption{Top:­ Time Series of 4 Her H$\alpha$ line from February to July 2009; Bottom: Time Series of 88 Her H$\alpha$ line from April to
July 2009. (Spectra are offset and labelled with the observation date, the oldest appears at the bottom and most recent at the top. Note that
although the spectra are displayed evenly spaced, they are not evenly distributed in time.)}
\label{Fig3}
\end{figure}

The material ejected from the central star forms an equatorially flattened disk also called the circumstellar decretion disk. We estimated the size of the disk by using the H$\alpha$ line . Figure \ref{Fig2} shows the H$\alpha$ line which is seen as double-peaked emission with the central absorption core between the two peaks going below the continuum for both the stars, as normally seen for any Be-shell star. The time series of the H$\alpha$ profile for both the stars is also shown in Figure \ref{Fig3}. 4 Her was observed 20 times in 11 days and 88 Her was observed 15 times in 8 days in the H$\alpha$ region.

The peak-to-continuum intensity \(I_p/I_c\) was obtained by considering the highest peak among the two peaks of H$\alpha$. De-blending technique in IRAF was used to estimate the equivalent width, EW of the H$\alpha$ line for all the days and is shown in separate tables for the two stars (Table~\ref{Tab6} and Table~\ref{Tab7}). 
%In the deblending technique, Gaussian profiles were fit for the emission peaks and Voigt profile was fit to the central absorption core. 
Another quantity measured from the spectra for the two stars is the velocity separation i.e., $\Delta$V between the red and violet peaks in the double-peaked profile. It will give an estimate about the region of emission of the H$\alpha$ profile. The violet-to-red peak intensity ratio measured with respect to continuum is called the V/R ratio. These were measured for 4 Her and 88 Her and all the quantities are shown in Table~\ref{Tab6} and Table~\ref{Tab7}. The average \(I_p/I_c\), EW, $\Delta$V and V/R values for 4 Her from 20 observations and 88 Her from 15 observations are shown in Table ~\ref{Tab8}. The error in EW and $\Delta$V are the standard deviation of the available observations.

\begin{table}
\centering
\caption{Measurements of H$\alpha$ emission line parameters for 4 Her.}
\begin{tabular}{ccccc}
\hline
\hline
$\bf Date\ of$ & \(I_p/I_c\) & $\bf EW$ & $\Delta$ $\bf V$ & $\bf V/R$\\
$\bf Observation$ & & $\bf ($\AA$) $ & ($\rm km $ $ s^{-1})$ & \\
\hline
03/02/09 & 1.47 & -7.0 & 280.0 & 1.16\\
04/02/09 & 1.44 & -6.4 & 281.0 & 1.16\\
26/04/09 & 1.64 & -8.6 & 257.7 & 1.03\\
26/04/09 & 1.64 & -7.5 & 274.3 & 1.01\\
27/04/09 & 1.62 & -7.4 & 260.7 & 1.03\\
27/04/09 & 1.61 & -7.2 & 281.6 & 1.06\\
28/04/09 & 1.69 & -7.2 & 262.8 & 1.11\\
28/04/09 & 1.58 & -6.8 & 262.0 & 1.07\\
29/04/09 & 1.73 & -8.4 & 274.4 & 1.13\\
29/04/09 & 1.66 & -7.2 & 253.0 & 1.15\\
30/04/09 & 1.64 & -7.2 & 253.9 & 1.10\\
30/04/09 & 1.63 & -6.9 & 256.7 & 1.13\\
01/05/09 & 1.64 & -7.6 & 274.0 & 1.17\\
01/05/09 & 1.64 & -7.4 & 280.6 & 1.18\\
11/07/09 & 1.62 & -8.0 & 275.3 & 0.94\\
11/07/09 & 1.62 & -7.4 & 300.1 & 0.94\\
22/07/09 & 1.69 & -8.2 & 292.0 & 0.91\\
22/07/09 & 1.68 & -7.8 & 281.3 & 0.94\\
23/07/09 & 1.60 & -7.3 & 273.0 & 0.86\\
23/07/09 & 1.69 & -8.3 & 284.7 & 0.96\\
\hline
\end{tabular}
\label{Tab6}
\end{table}

\begin{table}
\centering
\caption{Measurements of H$\alpha$ emission line parameters for 88 Her.}
\begin{tabular}{ccccc}
\hline
\hline
$\bf Date\ of$ & \(I_p/I_c\) & $\bf W$ & $\Delta$ $\bf V$ & $\bf V/R$\\
$\bf Observation$ & & $\bf ($\AA$) $ & ($\rm km $ $ s^{-1})$ & \\
\hline
26/04/09 & 2.16 & -11.8 & 316.9 & 0.85\\
26/04/09 & 2.11 & -10.2 & 313.4 & 0.82\\
27/04/09 & 2.21 & -12.0 & 317.6 & 0.83\\
28/04/09 & 2.03 & -9.8 & 298.8 & 0.91\\
28/04/09 & 2.05 & -9.2 & 318.6 & 0.85\\
29/04/09 & 2.11 & -10.8 & 319.4 & 0.88\\
29/04/09 & 2.02 & -9.4 & 318.7 & 0.87\\
30/04/09 & 2.06 & -9.0 & 326.0 & 0.86\\
30/04/09 & 2.04 & -9.5 & 318.1 & 0.85\\
01/05/09 & 2.07 & -10.2 & 318.4 & 0.90\\
01/05/09 & 2.01 & -10.5 & 301.0 & 0.93\\
01/05/09 & 2.05 & -9.8 & 296.1 & 0.88\\
11/07/09 & 1.93 & -11.2 & 297.8 & 0.96\\
11/07/09 & 1.95 & -13.4 & 297.3 & 0.96\\
27/07/09 & 2.12 & -11.4 & 312.4 & 0.93\\
\hline
\end{tabular}
\label{Tab7}
\end{table}

Rotational velocity of the disk is approximated by the power law as shown in Eq. \ref{eq2}. 
%mathematics mode
\begin{equation}
\frac{R_d}{R_*} = \left(\frac{2\ v\ sin\ \emph{i}}{\Delta V}\right)^\frac{1}{j} 
\label{eq2} 
\end{equation}

The extent of the H$\alpha$ emission region \( R_d\), is estimated in terms of the stellar radius \( R_*\) by assuming the region to be in Keplerian orbit around the star \citep{huang1972}. \( R_d\) is also estimated for non-Keplerian orbit by changing the rotational parameter \textit{j} from 1/2 to 1. \( R_d/R_*\) estimated for both the cases using the average $\Delta$V are shown in Table ~\ref{Tab8}. The $\textit{v}$ sin $\textit{i}$ is taken from Table ~\ref{Tab5}. 

We have not estimated the radius of the disk using \cite{catanzaro2013}, since the $\textit{v}$ sin $\textit{i}$ itself is too low. In the case of 4 Her, the radius of the disk is found to be in the range, 3.9 -- 5.6 \( R_*\) and for 88 Her it is found to be in the range, 3.4 -- 4.2 \( R_*\). \cite{slettebak1992} assumed Keplerian geometry for the circumstellar disk of 41 stars and concluded that the H$\alpha$ emission in general arises in the range 7 -- 19 \( R_*\) from the central star. Comparing our estimation of $\sim$ 5 \( R_*\) to this range, we can conclude that the H$\alpha$ emission arises closer to the star for both 4 Her and 88 Her and also that the H$\alpha$ emission region is smaller than the previously estimated average. We have not estimated errors in the values of the radius for each star, but present a range of values for them. We expect the errors to be about 10\%. The parameter which is susceptible to large errors is the equivalent width EW, where the error can be large when the continuum is not too well defined due to inadequate S/N ratio. 

\begin{table}
\centering
\caption{H$\alpha$ emission line parameters and estimation of the radius of the disk.}
\begin{tabular}{ccccccc}
\hline
\hline
$\bf Star$ & \(I_p/I_c\) & $\bf W$ & $\Delta$ $\bf V$ & $\bf \textit{v}\ sin\ \textit{i}$ \textsuperscript{\dag} & \multicolumn{2}{c}{ $\bf R$\begin{scriptsize}d\end{scriptsize}/$\bf R$\begin{scriptsize}*
\end{scriptsize}}\\
$\bf Name$ & & $\bf ($\AA$) $ & $(\rm km $ $ s^{-1})$ & $(\rm km $ $ s^{-1})$ & $\textit{j} = 1/2$ & $\textit{j} = 1$\\
\hline
 & & & & 310 & 5.19 & 2.28\\[0ex]
 & & & & 320 & 5.53 & 2.35\\[0ex]
 & & & & 321 & 5.57 & 2.36\\[0ex]
\raisebox{2ex}{4 Her} & \raisebox{2ex}{1.63} & \raisebox{2ex}{-7.48 $\pm$ 0.13} & \raisebox{2ex}{272.95 $\pm$ 2.9} & 269 & 3.91 & 1.98\\[0ex]
 & & & & 300 & 4.83 & 2.2\\[0ex]
 & & & & 275 & 4.06 & 2.02\\[0ex]
\hline 
 & & & & 316 & 4.13 & 2.03\\[0ex] 
 & & & & 303 & 3.43 & 1.85\\[0ex]
 \raisebox{0ex}{88 Her} & \raisebox{0ex}{2.06} & \raisebox{0ex}{-10.55 $\pm$ 0.32} & \raisebox{0ex}{311.36 $\pm$ 2.61} & 288 & 3.80 & 1.95\\[0ex]
 & & & & 300 & 3.71 & 1.93\\[0ex]
 & & & & 40 & --- & ---\\[0ex]
\hline
\multicolumn{7}{l}{\textsuperscript{\dag}\footnotesize{Refer Table~\ref{Tab5}}}\\
\end{tabular}
\label{Tab8}
\end{table}

\subsection{V/R variation}
\label{Sect4.3}

The material ejected from the central star moves around in the disk and when this system is viewed in a highly inclined or edge-on system, it results in the splitting of the H$\alpha$ line into two peaks. The H$\alpha$ emission line may be either single or double-peaked, but the most common profile seen are the double-peaked emission lines \citep{hanuschik1996}. The variation of the V/R ratio is called the V/R variation. The variation in the peaks can be seen clearly in the spectra for both the stars which is illustrated in Figure \ref{Fig3}. Both 4 Her and 88 Her were previously known to show V/R variations of 46.18 day \citep{koubsky1997} and 86.72 day \citep{doazan1982} period respectively. In this study, we looked for the short-term V/R variations in these two stars using different period search methods.

\cite{koubsky1997} used spectroscopic and photometric observations of 4 Her from 1969 -- 1997 which also included data from \cite{harmanec1976} . They  determined the period using the radial velocity measurements of the H$\alpha$ line. They reported a phase-locked V/R variation of period 46.18 days by considering two distinct shell episodes twenty years apart. \cite{rivinius2006} confirmed this period 
%\cite{koubsky1997}. \cite{rivinius2006} 
from their data set and reported a decrease of EW from -6 to -1~\AA ~from early 1997 to mid 1999 and again an increase to -5.5~\AA ~after which it remained unchanged till early 2003. In our observation, 4 Her shows H$\alpha$ emission with $V > R$ in all the spectra taken before July but showed a reversal i.e., $V < R$ for the three spectra obtained in the month of July. There is also a slight variation of EW for 4 Her from -6.4 to -8.6~\AA ~in a span of six months in 2009 which indicates that the EW has increased from the value quoted by \cite{rivinius2006}.

\cite{doazan1982} used 1963 -- 1979 spectroscopic data and 1968 -- 1981 photometric observations and determined the period for 88 Her as 86.7221 days and also reported that the period seems to be stable since 1912. They also reported that the V/R variation was in phase with the radial velocity of the H$\alpha$ line. \cite{duemmler1988} observed this star from 1977 -- 1987 and studied different lines in the shell spectrum. \cite{mennickent1991} reports that the V/R period of 88 Her is about 0.24yr and also gave the EW as 3.9~\AA ~during their observation before 1989. During our observation, 88 Her showed only slight fluctuations in the V/R ratio but had $V < R$ for all the spectra. \cite{rivinius2006} also reported $V < R$ in their observation. We see a significant variation of EW for 88 Her from -9.0 to -13.4~\AA ~in a span of four months in 2009. 

%\begin{figure}
%    \centering
%    \includegraphics[width=12cm,  height=12cm]{vbyr1.eps}    
%    \caption{V/R Variability of the double peaked H$\alpha$ emission line in 4 Her (Top) and 88 Her (Bottom) along with the over-plot of BeSS data shown in triangles.}
%    \label{Fig4}
%\end{figure} 

To search for the period of V/R variation in both the stars, we collected all the V/R values that were available in the literature and coupled it with our data to obtain robust results. For 4 Her, we collected V/R values from three previous data sets i.e., from \cite{harmanec1976} with 27 data points during 1969 -- 1973, from \cite{koubsky1997} with 89 data points during 1975 -- 1997 and \cite{rivinius2006} with 56 data points during 1998 -- 2003. We added 20 V/R values from our data set in 2009 which provided us altogether with 192 V/R data points for 4 Her. For 88 Her, we collected V/R values from two previous data sets i.e., from \cite{doazan1982} with 51 data points during 1971 -- 1979 and from \cite{duemmler1988} with 15 data points during 1981 -- 1987. We collectively had 81 V/R values along with our 15 data points for 88 Her.

\begin{figure}
    \centering
    \includegraphics[width=15cm,  height=13cm]{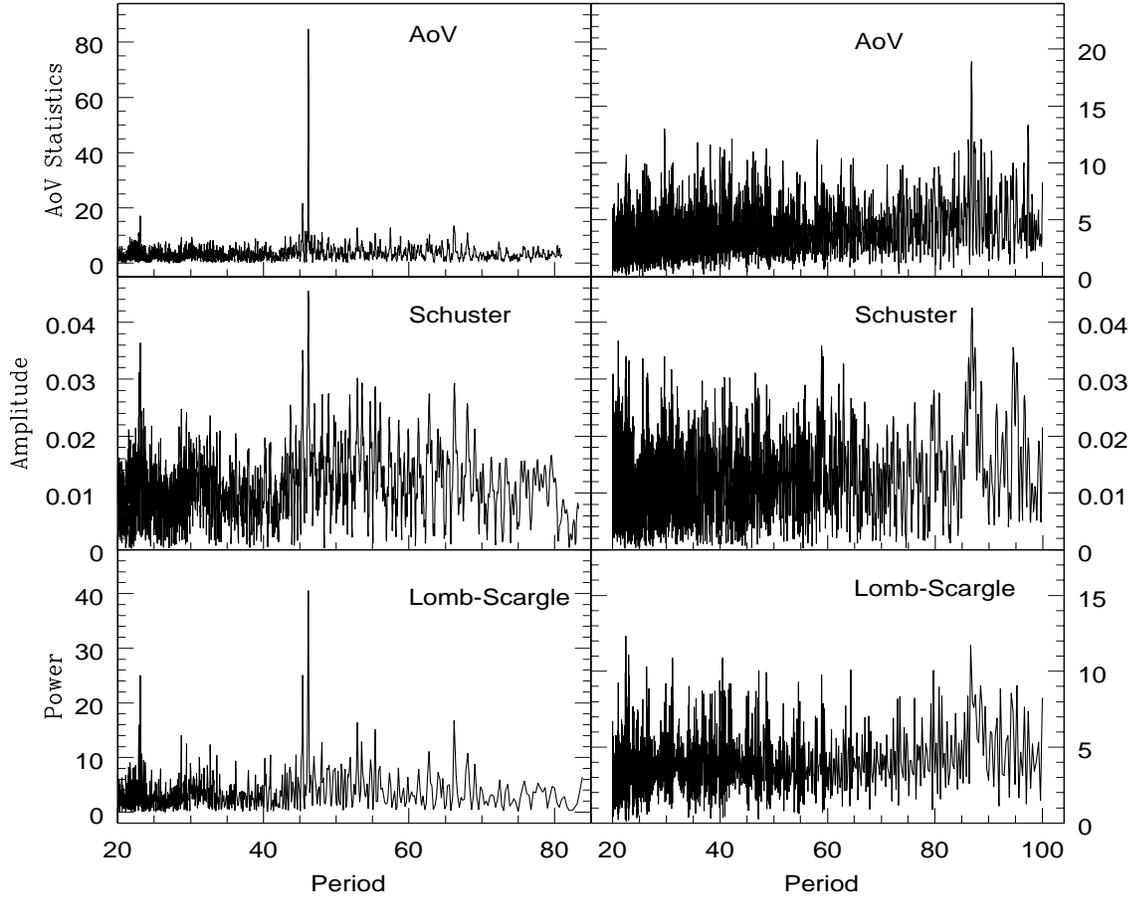}
    \caption{Periodograms for 4 Her (left panel) and 88 Her (right panel) using different period search algorithms.}
\label{Fig5}
\end{figure}

Two different approaches were adopted for the V/R period analysis, namely string-length and Fourier-based method. A time-series analysis has been performed with the rigorous analysis of variance (AoV) method in multi-harmonic (MAOVMH) mode \citep{schwarzenberg1996}. In this method, no implicit assumption is made on the shape of the variations. NASA Exoplanet Archive Periodogram Service with Lomb-Scargle algorithm \citep{scargle1982} was also used to estimate the Lomb-Scargle power spectra. In addition to this, we also used Schuster algorithm \citep{schuster1898} to derive the Schuster periodogram. In the latter case, we have estimated the Scargle False Alarm Probability (FAP) and Signal-to-noise ratio (S/N) of individual frequencies. The periods obtained from all the three methods are summarized in Table ~\ref{Tab9} for both the stars. Periodograms for both the stars using all the three methods are shown in Figure \ref{Fig5}.

\begin{table}
\centering
\caption{Periods obtained from different periodograms for 4 Her and 88 Her. Scargle False Alarm Probability (FAP) and Signal to Noise ratio of the peak in case of Schuster periodogram is given. p-value in the case of Lomb-Scargle algorithm is also given.}
\begin{tabular}{ccccccc}
\hline
\hline
$\bf Star\ Name$ & $\bf AoV$ & \multicolumn{3}{c}{$\bf Schuster$} & \multicolumn{2}{c}{$\bf Lomb-Scargle $} \\
\hline
          & Period (days) & Period (days) & FAP & S/N  & Period (days) & p-value      \\
\hline
4 Her     & 46.296        & 46.296        & 7.57811E-015 & 8.333 & 46.173        & 0            \\
          & 23.095        & 23.095        & 4.11814E-013 & 8.678 & 23.095        & 2.68922E-008 \\
\hline
88 Her    & 86.797        & 86.957        & 1.11085E-002 & 6.474 & 86.884        & 3.61152E-003 \\
\hline
\end{tabular}
\label{Tab9}
\end{table}

\begin{figure}
	\begin{subfigure}{0.5\textwidth}
    \centering
    \includegraphics[width=7cm,  height=6cm]{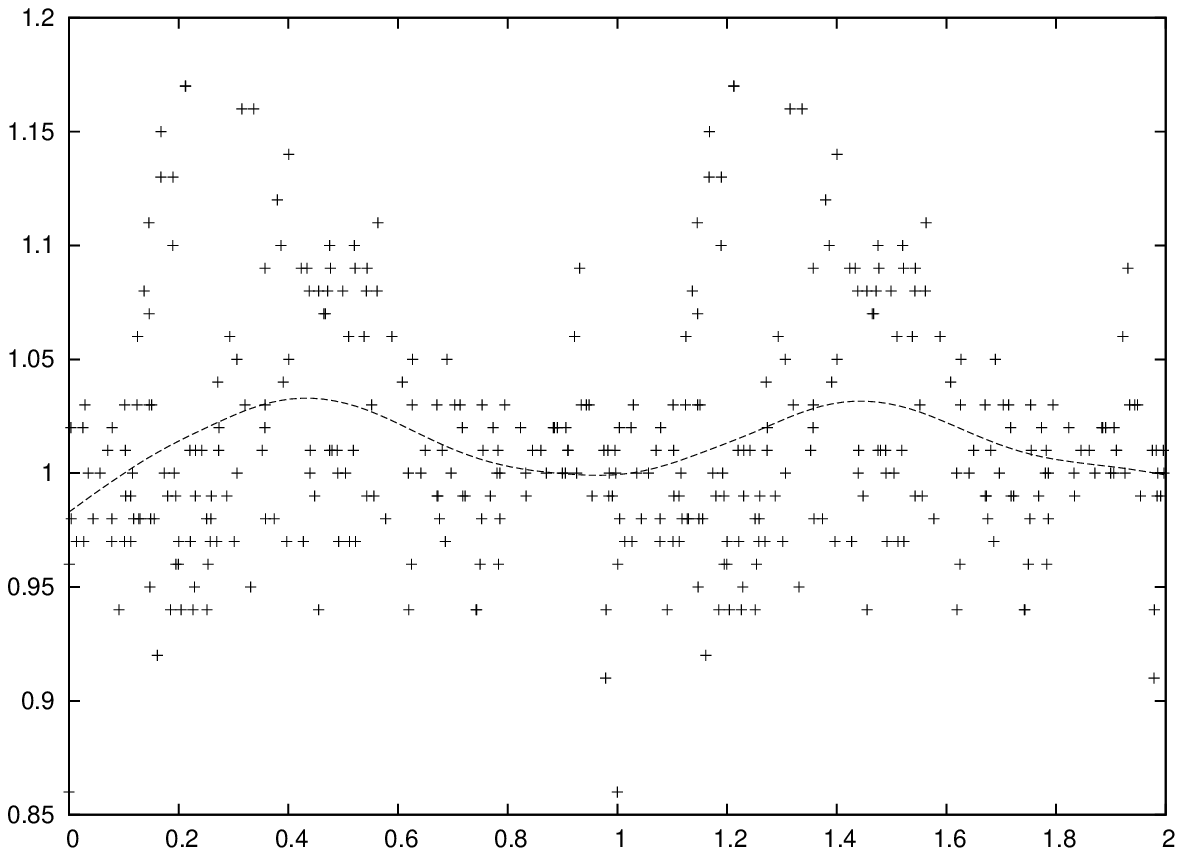}
    \end{subfigure}%
    %\hspace{0.3cm}
    \begin{subfigure}{0.5\textwidth}
    \centering
    \includegraphics[width=7cm,  height=6cm]{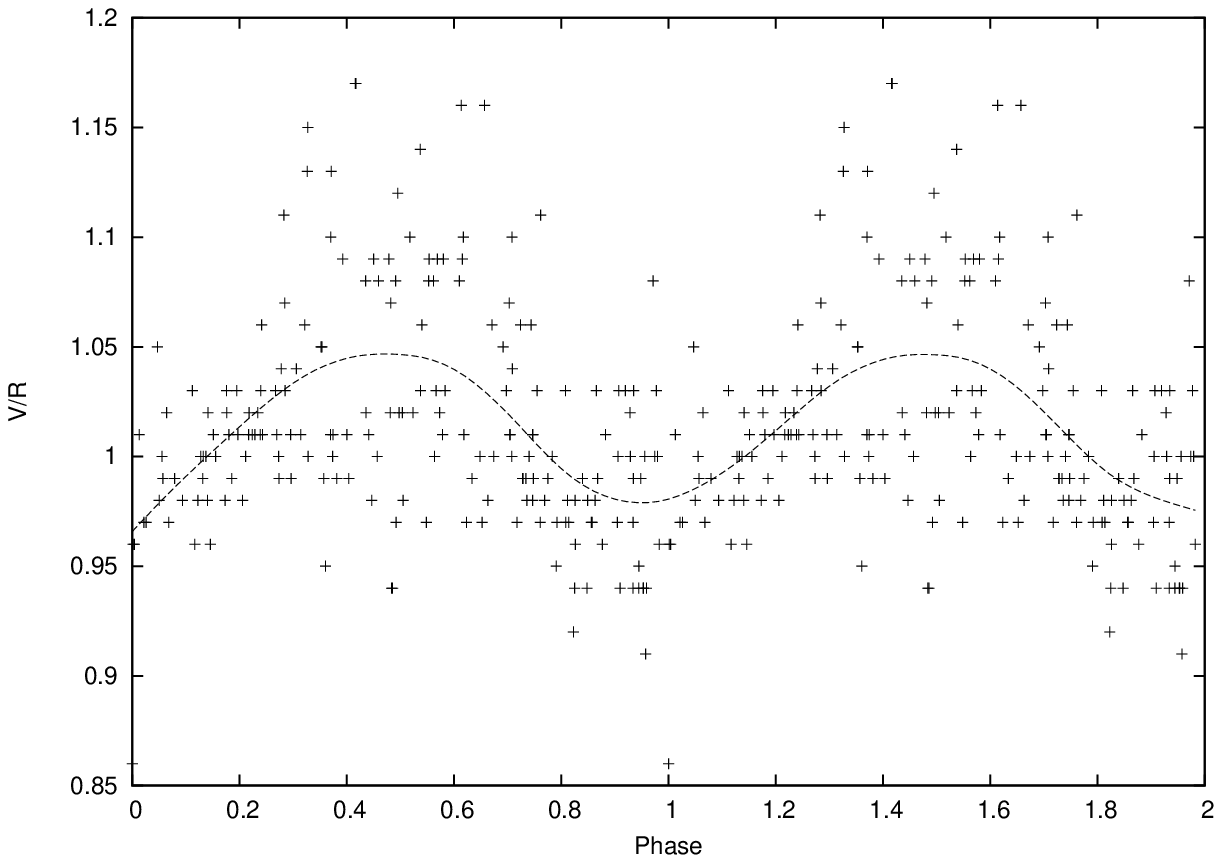}
    \end{subfigure}%    
    \caption{Phase plot of V/R for 4 Her with period 46.296 days (left) and 23.095 days (right) along with a fit of aspline function.}
\label{Fig4}
\end{figure}

All the three methods yielded almost the same frequency spectrum (Fig \ref{Fig5}) for both the stars. In addition to the longer $\sim$ 46 day period of 4 Her, we detected a smaller period of 23.095 days which is the harmonic of the $\sim$ 46 day period. Agreement between the result of different methods was used as an indication of reliability of a detection. In order to check whether the shorter period was already there in the older data set, we checked  in the data sets from \cite{harmanec1976}, \cite{koubsky1997} and \cite{rivinius2006}. We obtained the above mentioned shorter period in the data set from \cite{koubsky1997} which had 89 V/R values, but it was not as significant as what we obtained after combining all the data and our data. This was also checked by combining the other data sets to these 89 V/R values, where we observed an increasing trend of the significance of the period with the addition of more data. Thus, the new period we estimated might have been missed earlier due to sparse data set. We conclude that the V/R variation in 4 Her is more than singly periodic. But without having a larger dataset available, further discussions on multiple periodicity would seem meaningless. This suggests that continuous monitoring of this star is necessary to obtain the V/R variability period. A phase plot of V/R has been shown in Figure \ref{Fig4} for 4 Her with two periods detected. No smaller periods were detected in case of 88 Her except the already known $\sim$ 86 day period. The phase plot for 88 Her was not included since the data set is very sparse. However, more continuous observations are required to understand its nature of variability. We discuss the implication of this result in the next section.

\subsection{Discussion}
\label{Sect4.4}

In this study, we have estimated various parameters of the two shell stars, 4 Her and 88 Her. We estimated the rotational velocity, $\textit{v}$ sin $\textit{i}$ of the stars to be 305 $\pm$ 13$\rm km $ $ s^{-1}$ for 4 Her and 302 $\pm$ 16$\rm km $ $ s^{-1}$ for 88 Her. These stars are well known to be edge on systems, hence these velocities are considered very close to the true velocity. By assuming the critical velocity based on the spectral class, we estimated that the fractional critical rotation is about 0.8, suggesting that these stars are rotating very close to the breakup velocity. We also estimated the radius of H$\alpha$ emission, and is found to be in the range, 3.4 -- 5.6 \( R_*\). In summary, we find that these stars are fast rotating stars with the H$\alpha$ emission region located very close to the star. 

We also notice that there is a V/R variation in the H$\alpha$ profile. V/R variation occurs only when the disk is assumed to be an eccentric Keplerian disk. This is understood as a global density wave pattern in the circumstellar disk as shown in Figure 2 of \cite{McDavid2000} of a one-arm density wave for $\zeta$ Tau. It is the precession of the density wave about the central star.  This is based on the model of global one-armed oscillations of equatorial disks in Be stars as given by \cite{okazaki1991}. 

4 Her is reported to have V/R variations phase-locked to the orbit of the binary component \citep{stefl2007}. \cite{rivinius2006} compared the V/R variations of 4 Her and $\epsilon$ Cap with $\phi$ Per and mentions that the variations are periodic for the phase-locked V/R variation rather than being cyclic and also observed that they have shorter time-scales. \cite{mennickent1991} found that 6 among their sample of 33 stars, showed short term V/R variations and these are stars with small envelopes. They suggest that the short term V/R variations could be caused by the rotation of inhomogeneities in the circumstellar envelopes. As the two shell stars studied are binaries, it is quite possible that the inhomogeneities in the circumstellar material is caused by the gravitational effect of the binary. This is also supported by the fact that one of the observed period of 4 Her is a harmonic of the previously estimated period. Thus, the change in the value of the period estimated suggests that there might be an inhomogeneity present in the disk and probably the binary star may be responsible for this perturbation.

Further observations and continuous monitoring of these stars will increase the number of data points and help us in identifying the period more accurately. Our study also reveals that short-term monitoring of these systems is important to understand the effect of the binary star and its perturbation on the circumstellar disk. This study finds that 4 Her and 88 Her are ideal targets for continuous monitoring of V/R variation in the H$\alpha$ profile and especially since these observations can be performed with moderate telescope equipped with a spectrograph.

\section{Conclusions}
\label{Sect5}
\begin{enumerate}
\item We have presented the spectroscopic analysis of the two Be-shell stars, 4 Her and 88 Her which was observed for about six months in 2009.
\item The rotational velocity, $\textit{v}$ sin $\textit{i}$ was calculated using He{\sc i} lines and is found to be $\sim$ 300$\rm km $ $ s^{-1}$ for both the stars. The fraction of critical rotation for the two stars is found to be $\sim$ 0.8 suggesting them to be a rapid rotators.
\item The radius of the circumstellar disk \( R_d/R_*\) using the H$\alpha$ double-peaked emission profile is found to be $\sim$ 5.0, assuming a Keplerian orbit for both 4 Her and 8 Her. This implies that the H$\alpha$ emission disk is very small for both the stars.
\item The EW of the H$\alpha$ emission line profile varied from -6.4 to -8.6~\AA ~for 4 Her in a span of six months and from -9.0 to -13.4~\AA ~for 88 Her in a span of four months.
\item V/R variation was observed for both stars and the period was re-estimated by different period search techniques. For 4 Her, two periods $\sim$ 46 days and 23.095 days, a harmonic of each other was detected. For 88 Her, longer period of $\sim$ 86 days, which is very close to the literature period was obtained; but further confirmation is required after continuous monitoring.
\item We conclude and confirm that these two stars are rapid rotators with smaller H$\alpha$ emitting region. As they have binaries (with 4 Her being a phase locked binary), the observed V/R variation may be due to the gravitational effect of the binary on the circumstellar disk. 
\end{enumerate}

\normalem
\begin{acknowledgements}
This work was funded by Centre for Research, Christ University, Bangalore as a part of Major Research Project. We thank the support staff at 1.0m telescope, Vainu Bappu Observatory, especially Jayakumar K., for their assistance in obtaining the data used in this paper. This research has made use of the NASA Exoplanet Archive, which is operated by the California Institute of Technology, under contract with the National Aeronautics and Space Administration under the Exoplanet Exploration Program. The first author would also like to thank Sowgata Chowdhury from Christ University for assistance with period search algorithms. 
\end{acknowledgements}

\bibliographystyle{raa}
\bibliography{ref}

\end{document}